\documentclass[aps,prd,preprint,preprintnumbers,showpacs]{revtex4-1}
\usepackage{graphicx}
\usepackage{amsmath}
\usepackage{relsize}
\usepackage{caption}
\usepackage[section] {placeins}
\usepackage{slashed}
\usepackage[dvips]{color}
\usepackage{soul}
\usepackage{changepage}
\begin{document}

\title{Relating   $D^* \bar{D}^*$ currents with $J^\pi= 0^+,1^+$ and $2^+$ to $Z_c$ states }
\author{K.~P.~Khemchandani\footnote{kanchan@if.usp.br}}
\author{A.~Mart\'inez~Torres\footnote{amartine@if.usp.br}}
\author{ M.~Nielsen\footnote{mnielsen@if.usp.br} }
\author{ F.~S.~Navarra\footnote{navarra@if.usp.br}}
\preprint{}

 \affiliation{
Instituto de F\'isica, Universidade de S\~ao Paulo, C.P 66318, 05314-970 S\~ao Paulo, SP, Brazil.}

\date{\today}

 \begin{abstract}
In this work we study the $D^* \bar{D}^*$ current with QCD sum rules. We write the correlation function using the general current corresponding to the $D^* \bar{D}^*$ system 
and then use spin projectors to obtain the correlation function in the $0^+, 1^+$ and $2^+$ spin-parity configurations. The purpose of the present work is to investigate the possibility of explaining  the recently reported $Z_c(4025)$ as a $D^* \bar{D}^*$ moleculelike state.  As a result we find a state for each spin case with a very similar mass:  $M^{S=0} = \left(3943 \pm 104\right)$ MeV, $M^{S=1} = \left(3950 \pm 105\right)$ MeV, $M^{S=2} = \left( 3946 \pm 104 \right)$ MeV.  We discuss that our mass results, within error bars, for  $1^+$ or  $2^+$ are both  compatible with   $Z_c(4025)$. However, our results are also compatible with a possible $D^* \bar{D}^*$ bound state, in agreement with  predictions of some previous works. We have also calculated the current-state coupling which turns out to be larger in the $2^+$ case. 
\end{abstract}

\maketitle

\section{Introduction}
The present work has been motivated by the finding of a charged charmoniumlike state, named $Z_c^\pm(4025)$, in the $e^+ e^- \to \left( D^* \bar{D}^* \right)^\pm \pi^\pm$ process by the BES collaboration \cite{bes4025}.  The state is found in the pion recoil mass spectrum  (which corresponds to the $D^* \bar{D}^*$ invariant mass spectrum)  at $4026.3 \pm 2.6 \pm 3.7$ MeV,  very close  to the $D^* \bar{D}^*$  threshold.  The spin-parity of  $Z_c^\pm(4025)$ is not known although it has been assumed to be $1^+$ in Ref.~\cite{bes4025}.  Its isposin is obviously 1. Interestingly, another  state with very similar mass, $Z_c(4020)$ has also been found in the $\pi^\pm h_c$ mass spectrum in the $e^+ e^- \to \pi^+  \pi^- h_c$ process \cite{bes4025_2} although it does not seem to be clear  if the states found in $\left( D^* \bar{D}^* \right)^\pm$ and $\pi^\pm h_c$ are the same or not. 

The finding of these new $Z_c$ states adds to the discovery of a series of bottomoniumlike/ charmoniumlike charged states which are being reported from recent experimental studies. The special interest in these states  arises from the fact that they necessarily require more than two constituent quarks to get  their quantum numbers right.  The existence of  mesons (baryons) possessing  more than two (three) constituent  quarks has always been viable within QCD. However, a rigorous experimental search for such states has begun only recently, since the states made of heavy quarks are now produced with high statistics in the new $B$ factories. This makes it  easier to look for exotic   hadrons by searching for  charged heavy quarkoniumlike states. 

In the charm sector, several charged states have been reported by now, like $Z^+ (4430)$, found in the $\pi^+ \psi^\prime$ system \cite{belle4430_1,belle4430_2} (also reconfirmed in Ref.~\cite{belle4430_3} more recently), $Z_1^+ (4050)$, $Z_2^+(4250)$ found in the $\pi^+ \chi_{c1}$ invariant mass spectrum \cite{belle4050_4250}, $Z_c (3900)$ in the $\pi^\pm J/\psi$ system \cite{bes3900,belle3900} and now $Z_c^\pm(4025), Z_c(4020)$ found in the $D^* \bar{D}^*$  \cite{bes4025} and $\pi h_{c1}$ \cite{bes4025_2} systems, respectively. It should be mentioned, however, that the first three of these $Z_c$ states have not been found by the Babar collaboration \cite{babar4430,babar4050_4250}. Hence, some of these states still need more confirmation for their existence. But if they do exist then it is interesting to notice that the masses of all these states are relatively close to each other, while the widths of these states vary between 40-200 MeV. Curiously, all of them lie very close to the threshold  of some open charm meson system. Also, all the decay channels where these charged states have been found (like, $\pi^+ \psi^\prime$,  $\pi^+ \chi_{c1}$, $\pi^+ J/\psi$, $D^*\bar{D}^*$) can account for similar isospin-spin quantum numbers. Thus, these channels, in principle, can couple to each other and  it is possible that a same $Z_c$ state is seen in different $c\bar{c}-\pi$ or open charm final states. In such a situation, where  states  with closely spaced masses and overlapping widths   are being found, it is very important to make a careful analysis to judge if all of them are different or are sometimes replicas of each other. 

To add to the efforts in understanding these newly found states, we make a  study of the $D^* \bar{D}^*$ molecule-like current in the isospin 1 configuration using QCD sum rules and study its different spin configurations. 

A system of two vector mesons can possess a total spin-parity $0^+$, $1^+$ or $2^+$ when interacting in s-wave. Such configurations of two vector mesons are ideally suited to formation of moleculelike resonances as the constituent hadrons posses little energy.  Since the masses of  some $Z_c$'s are close to the threshold of open charm meson systems, some of them could be explained  within such a picture. To unambiguously  separate the different spin-parity configurations of the $D^* \bar{D}^*$ systems, we apply the spin projectors discussed in our previous work \cite{us} on the most general  current written for the system. Some works have already been done on the  $D^* \bar{D}^*$ system using QCD sum rules with the motivation of finding a state which can be associated to $Z_c(4025)$.  In Refs.~\cite{Chen:2013omd,Cui:2013vfa} a current corresponding to $1^+$ spin-parity has been studied and a state compatible with $Z_c(4025)$ has been found. The authors of Ref.~\cite{Qiao:2013dda} investigate a tetraquark  current with spin-parity $1^-$ and $2^+$ to conclude that $Z_c(4025)$ is a $2^+$ tetraquark state. Some work has also been done with other formalisms to understand the nature of $Z_c(4025)$  \cite{He:2013nwa,Wang:2013qwa,Guo:2013sya}. The $D^* \bar{D}^*$ system has also been studied in Ref.~\cite{raquel2} where a  bound state in isospin 1  and spin-parity $2^+$ was predicted with mass between 3900-3965 MeV and width of 160-200 MeV. We will later compare our results with those found in these previous works and make some conclusions.

\section{Formalism}
We write the interpolating current corresponding to the  $\bar{D}^{*0} D^{*+}$ molecule as
\begin{equation}\label{j}
j_{\mu \nu} (x) = \left[ \bar{c}_a(x) \gamma_\mu u_a(x)\right]\left[\bar{d}_b(x) \gamma_\nu c_b(x)\right],
\end{equation}
where $a,b$ denote the color indices. With this current 
 we construct the two-point  correlation function
\begin{equation}
\Pi_{\mu \nu \alpha \beta} (q^2) = i \int  d^4x e^{iqx} \langle 0 \mid T \left[ j_{\mu \nu} (x) j^\dagger_{\alpha \beta} (0) \right] \mid 0 \rangle \label{Pi}
\end{equation}
and then apply the spin projectors discussed in Ref.~\cite{us} to it. The $0^+$, $1^+$ and $2^+$ components of the correlation function written in Eq.~(\ref{Pi}) can be 
obtained using the following projectors
\begin{align} 
\mathcal{P}^{(0)}&=\frac{1}{3}\Delta^{\mu\nu}\Delta^{\alpha\beta},\nonumber\\
\mathcal{P}^{(1)}&=\frac{1}{2}\left(\Delta^{\mu\alpha}\Delta^{\nu\beta}-\Delta^{\mu\beta}\Delta^{\nu\alpha}\right),\label{proj}\\
\mathcal{P}^{(2)}&=\frac{1}{2}\left(\Delta^{\mu\alpha}\Delta^{\nu\beta}+\Delta^{\mu\beta}\Delta^{\nu\alpha}\right)-\frac{1}{3}\Delta^{\mu\nu}\Delta^{\alpha\beta},\nonumber
\end{align}
where $\Delta_{\mu\nu}$ is defined in terms of the metric tensor, $g^{\mu\nu}$, and the four momentum $q$ of the correlation function as
 \begin{align}
\Delta_{\mu\nu}\equiv -g_{\mu\nu}+\frac{q_\mu q_\nu}{q^2}.\label{Delta}
 \end{align}
These projectors were obtained in  Ref.~\cite{us} by building an analogy with the work done in Ref.~\cite{raquel} where  the s-wave $D^* \rho$ interaction was studied using effective field theory. Some of these projectors coincide with those determined in  Ref.~\cite{project}, where projectors for more spin-parity cases  are given. As mentioned earlier, we are interested in studying $0^+$, $1^+$ and $2^+$ configurations of $D^* \bar{D}^{*}$ keeping in mind that the low energy interaction of these two mesons is dominated by s-wave scattering which is a favorable situation for formation of moleculelike states.

The motivation behind separating only the positive parity components is to look for moleculelike states with mass close to the threshold of the constituent hadrons, in which case there is little energy available for the hadrons, which as a consequence interact in s-wave.  
 A moleculelike  picture for  $Z_c(4025)$  seems to be quite plausible since its mass is merely 8 MeV away from  the $\bar{D}^{*0} D^{*+}$ threshold. In other words, here we want to see if $Z_c(4025)$ can be interpreted as a  $1^+$ or $2^+$ resonance of the  $\bar{D}^{*0} D^{*+}$ system. The $0^+$ assignment is ruled out for $Z_c(4025)$ by spin-parity conservation for the
  $e^+ e^- \to \left( D^* \bar{D}^* \right)^\pm \pi^\pm$ process.  However,  some other $Z_c$ resonance with  $0^+$ might exist.

As is well known, the QCD sum rules method is based on the dual nature of the correlation function: it can be interpreted as quark-antiquark fluctuations at short distances, which is usually referred to as the QCD side, while it can be related to hadrons at large distances, which is  referred to as the phenomenological side.  In this method, thus one calculates the correlation function within both interpretations and  equates the two results with the conviction that the two sides must be equivalent in some range of $q^2$ \cite{svz,reviews1,reviews2,reviews3}. The calculation from the QCD side leads to a quark propagator form of the  correlation function which is written in terms of the  operator product expansion (OPE)  and  the coefficients of the series are calculated  perturbatively \cite{svz,reviews1,reviews2,reviews3,reviews4}.

In practice, one calculates the spectral density which is related to the correlation function through the dispersion relation
\begin{align}
 \Pi_{\textrm{OPE}}(q^2)=\mathlarger{\int}\limits_{smin}^\infty ds \,\, \frac{\rho_{\textrm{OPE}}(s)}{s-q^2} + {\rm subtraction\,\, terms}.\label{corrope}
\end{align}
Proceeding with the standard scheme, then, we obtain the spectral density, corresponding to the spin-projected correlation function,  by going in the OPE series up to dimension six in the present case
\begin{equation}
\rho^{S}_{\textrm{OPE}}=\rho^{S}_{\textrm{pert}}+\rho^{S}_{\langle\bar q q\rangle}+\rho^{S}_{\langle g^2 G^2\rangle}+\rho^{S}_{\langle\bar q g\sigma G q\rangle}+\rho^{S}_{{\langle\bar q q\rangle}^2}+\rho^{S}_{\langle g^3 G^3\rangle}.\label{rho}
\end{equation}
The different spin-projected OPE results are given in the appendix  of the paper.

Next, to make the calculations from the  phenomenological side we assume, as usually done, that the spectral density can be written as a sum of a narrow, sharp state, which precisely corresponds to the one we are looking for, and a smooth continuum 
\begin{align}
\rho^{S}_{\textrm{phenom}}(s)={\lambda^2_{S}}\delta(s-m_{S}^2)+\rho^{S}_{\textrm{cont}}(s). \label{rhopheno}
\end{align}
In Eq.~(\ref{rhopheno}) $S$ denotes the spin, $s = q^2$ is the squared four-momentum flowing in the correlation function,  $\lambda_{S}$ is the coupling of the current to the state we are interested in and $m_{S}$ denotes its mass.
The density related to the continuum of states is assumed to vanish below a certain value of $s$ called continuum threshold, let us call it $s_0$. Above $s_0$ one usually  considers the ansatz~\cite{svz,reviews1,reviews2,reviews3}
\begin{align}
\rho_{\textrm{cont}}(s)=\rho^{S}_{\textrm{OPE}}(s)\Theta(s-s_0).\label{ansatz}
\end{align}
Using this parametrization of the spectral density, the correlation function from the phenomenological side can be written as 
\begin{align}
 \Pi^{S}_{\textrm{phenom}}(q^2)=\frac{\lambda^2_{S}}{m_{S}^2-q^2}+\int_{s_0}^\infty ds\,\frac{\rho^{S}_{\textrm{OPE}}(s)}{s-q^2}.\label{corrphen}
\end{align}
To get closer to the  idea of the dual nature of the correlation function, a Borel transform of  Eqs.~(\ref{corrope}) and (\ref{corrphen}) is taken.  This suppresses  the contribution of the continuum on the phenomenological side and divergent contributions arising due to the long range interactions on the OPE side. Equating the Borel transformed results, we get the following expression for the mass 
\begin{align}
m_{S}^2=\frac{\int_{4 m^2_c}^{s^{S}_0}ds\,s \rho^{S}_{\textrm{OPE}}(s)e^{-s/M^2}}{\int_{m^2_c}^{s_0}ds\,\rho^{S}_{\textrm{OPE}}(s)e^{-s/M^2}},\label{mass}
\end{align}
and that for  the coupling $\lambda_{S}$ 
\begin{align}
\lambda^2_{S}=\frac{\int_{4 m^2_c}^{s_0}ds\,\rho^{S}_{\textrm{OPE}}(s)e^{-s/M^2}}{e^{-m_{S}^2/M^2}}.\label{lambda}
\end{align}

\section{Results and discussions}
In this article we are investigating the possibility of interpreting   some  of the recently found $Z_c$ states, which essentially need
more than two valence quarks, with a $D^* \bar{D}^*$ molecular current. For this, as explained in the previous section, we use the QCD sum rules method for which we write the general
current corresponding to $D^* \bar{D}^*$  system (Eq.~(\ref{j})) and then use spin projectors (Eq.~(\ref{proj})) to obtain the correlation function 
with spin-parity $0^+$, $1^+$ and $2^+$. To obtain the results, we need  first  to find a valid ``Borel window"  in which the results can be relied upon.

A valid Borel window is that range of Borel mass  where, on the QCD side, the OPE series converges and where the contribution from the pole term dominates over the one of the continuum on  the phenomenological side.  In order to carry out these calculations we need to fix the value of the continuum threshold, $\sqrt{s_0}$, which should be a reasonable value above the mass of the state we are interested in. Usually it is taken  to be 0.5 GeV above the mass of the state since, phenomenologically,  the average difference between the masses of a hadron and its first excited state is found to be around 0.5 GeV. Since the principal motivation of our work is to find a description for the recently found  $Z_c(4025)$ \cite{bes4025}, we take $\sqrt{s_0} \sim 4.5$ GeV. We will, actually, vary the value of $\sqrt{s_0}$ around 4.5 GeV and test the stability of our results against the variation of this value. We find that a valid Borel window exists for the calculations done with the  three spin-projected  correlation functions. It remains to give the values of the other inputs required for the numerical calculations, like the quark condensate, the gluon condensate, the constituent charm quark mass, etc. We use the same values for these inputs as those used in our previous work~\cite{our1}. For the readers convenience we also list them here in Table~\ref{parameters}.
\begin{table}[h!]
\caption{Values of the different inputs required for numerical
calculations.}\label{parameters}
\begin{ruledtabular}
\begin{tabular}{cc}
Parameters & Values\\
\hline
$m_c$ & $1.23 \pm 0.05$ GeV\\
$\langle \bar{q} q \rangle$ &$-(0.23 \pm 0.03)^3$ GeV$^3$\\
$\langle g^2 G^2 \rangle$ &$(0.88\pm0.25)$ GeV$^4$\\
$\langle g^3 G^3 \rangle$ & $(0.58\pm0.18)$ GeV$^6$\\
$\langle \bar{q} \sigma \cdot G q \rangle$& 0.8$\langle \bar{q} q \rangle$ GeV$^2$\\
\end{tabular}
\end{ruledtabular}
\end{table}

\begin{figure}[t]
\includegraphics[width= 0.49\textwidth, height=8cm]{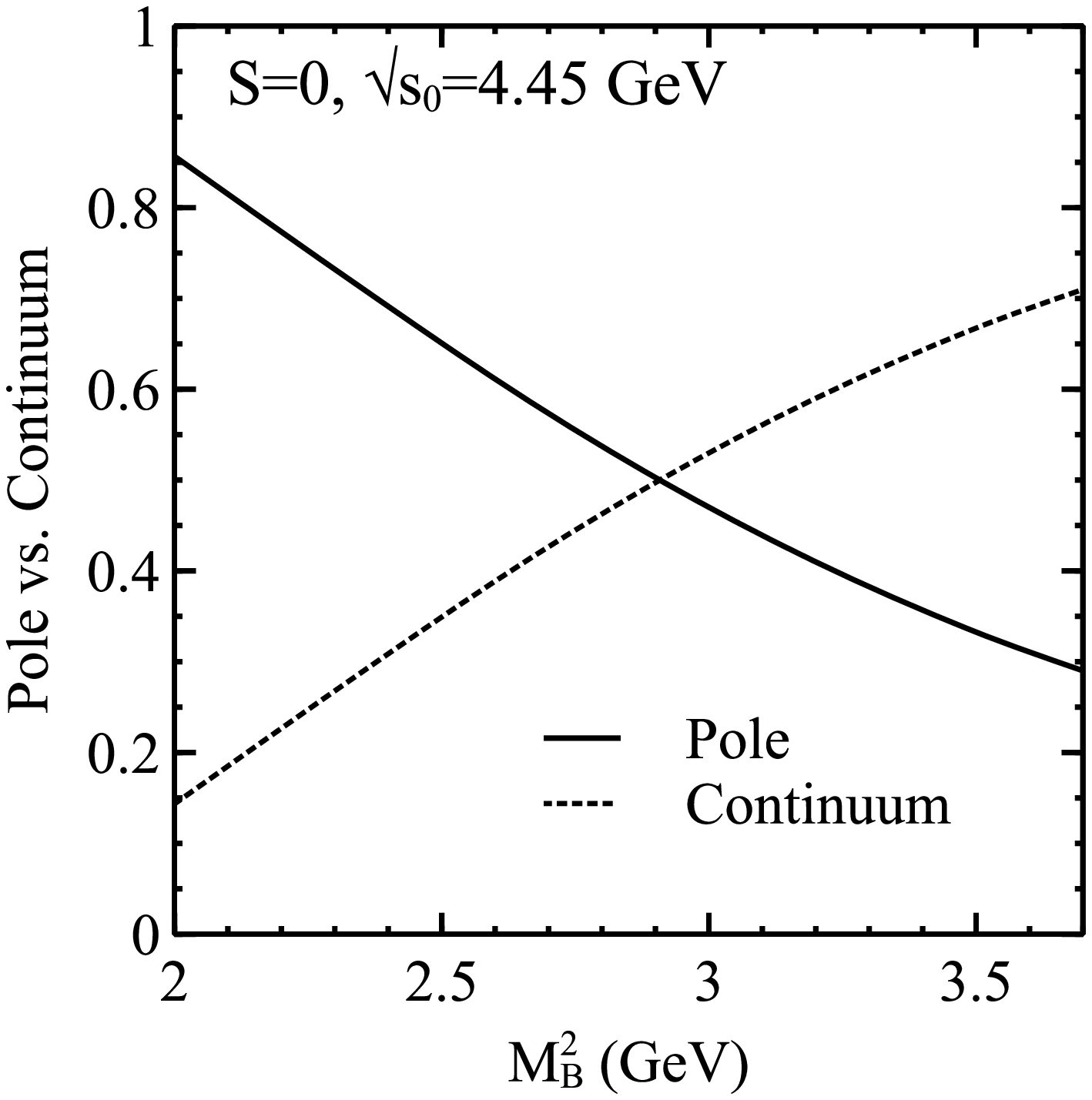} \includegraphics[width= 0.49\textwidth, height=8cm]{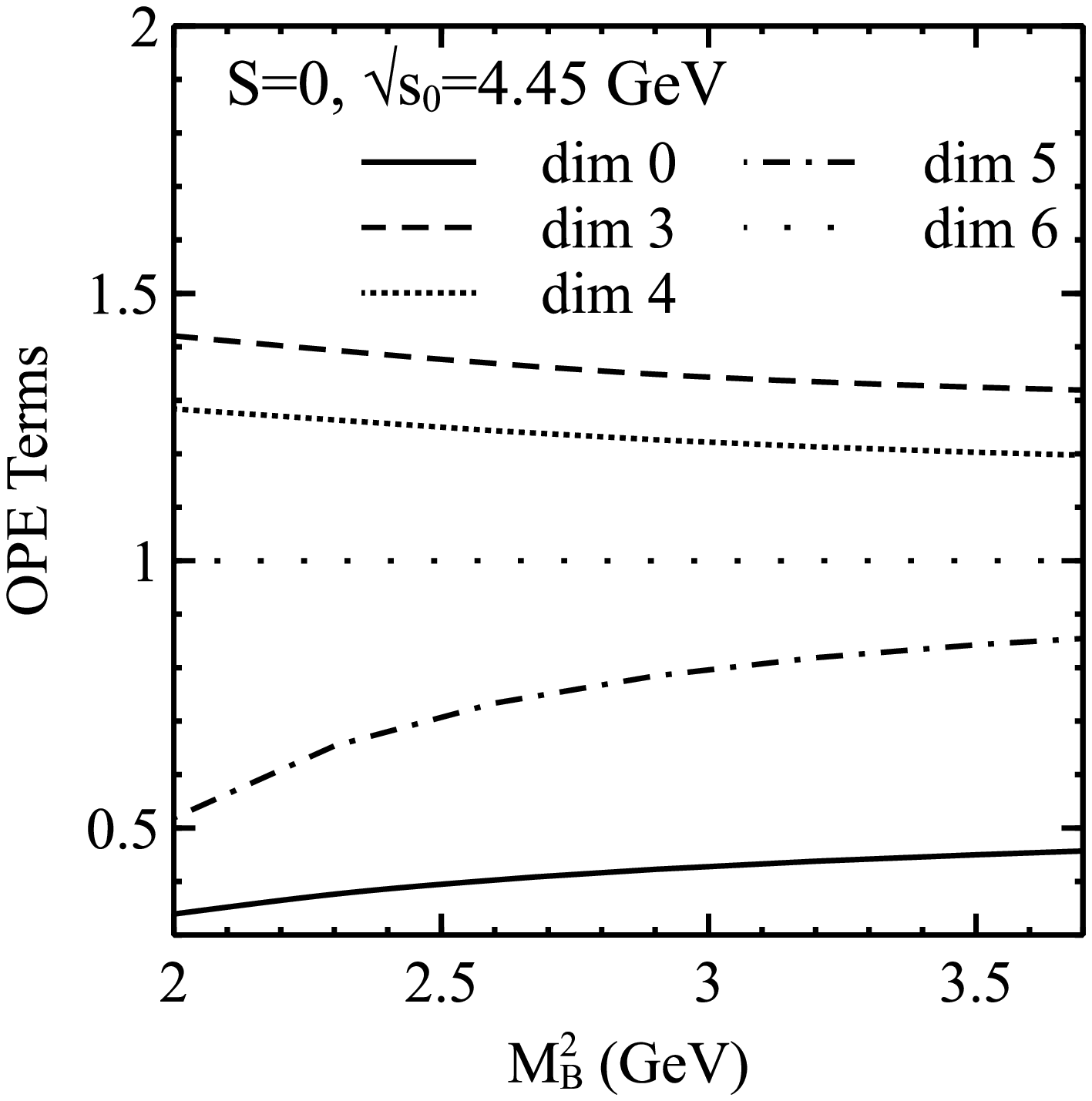}
\caption{Left panel: the  pole and continuum contributions for the correlation function corresponding to the $D^* \bar{D}^*$ system with $J^\pi = 0^+$. Right panel: contribution of the different terms of the OPE series for the same. The label $S$ in the figure denotes the spin (=0) of the system, while $\sqrt{s}_0$ denotes the continuum threshold which is taken as 4.45 GeV to obtain the results shown here. $M_B^2$ is the squared Borel mass. }
\label{polecont}
\end{figure}

In the left panel of  Fig.~\ref{polecont} we show the  contributions of the pole and continuum terms obtained by  calculating the correlation function from the phenomenological side. This figure shows the results for the correlation function for spin 0 and  for $\sqrt{s_0} \sim 4.45$ GeV, as an example. The results for other configurations and other values of  $\sqrt{s_0}$ around 4.5 GeV are similar.  The right panel shows the results obtained for the different terms in the OPE series for the same spin and $\sqrt{s_0}$. The OPE results shown in Fig.~\ref{polecont} are the relative contributions of the different terms of the series. This means that the result for dimension 0 (labeled by ``dim 0" in Fig.~\ref{polecont}) is divided by  the sum of all the terms in the series of Eq.~(\ref{rho}). Then the dimension 3 results are added to dimension 0 and the result is, once again, divided by the sum of all the terms in Eq.~(\ref{rho}) (labeled by ``dim 3" in Fig.~\ref{polecont}). Similarly, one keeps going to the next higher dimension. Thus the legend labels in  Fig.~\ref{polecont} indicate the dimension up to which the OPE terms, weighted by the sum of  all terms in Eq.~(\ref{rho}), have been considered.

 As can be seen from Fig.~\ref{polecont}, the pole term,  on the phenomenological side, dominates up to a squared Borel mass $\sim$ 2.9 GeV$^2$ while the convergence of the OPE series is good beyond the squared Borel mass $\sim$ 2.65 GeV$^2$ (where the contribution of the second last term in the series is $\le$25$\%$ of the last term). 

\begin{figure}
\includegraphics[width= 0.49\textwidth, height=8cm]{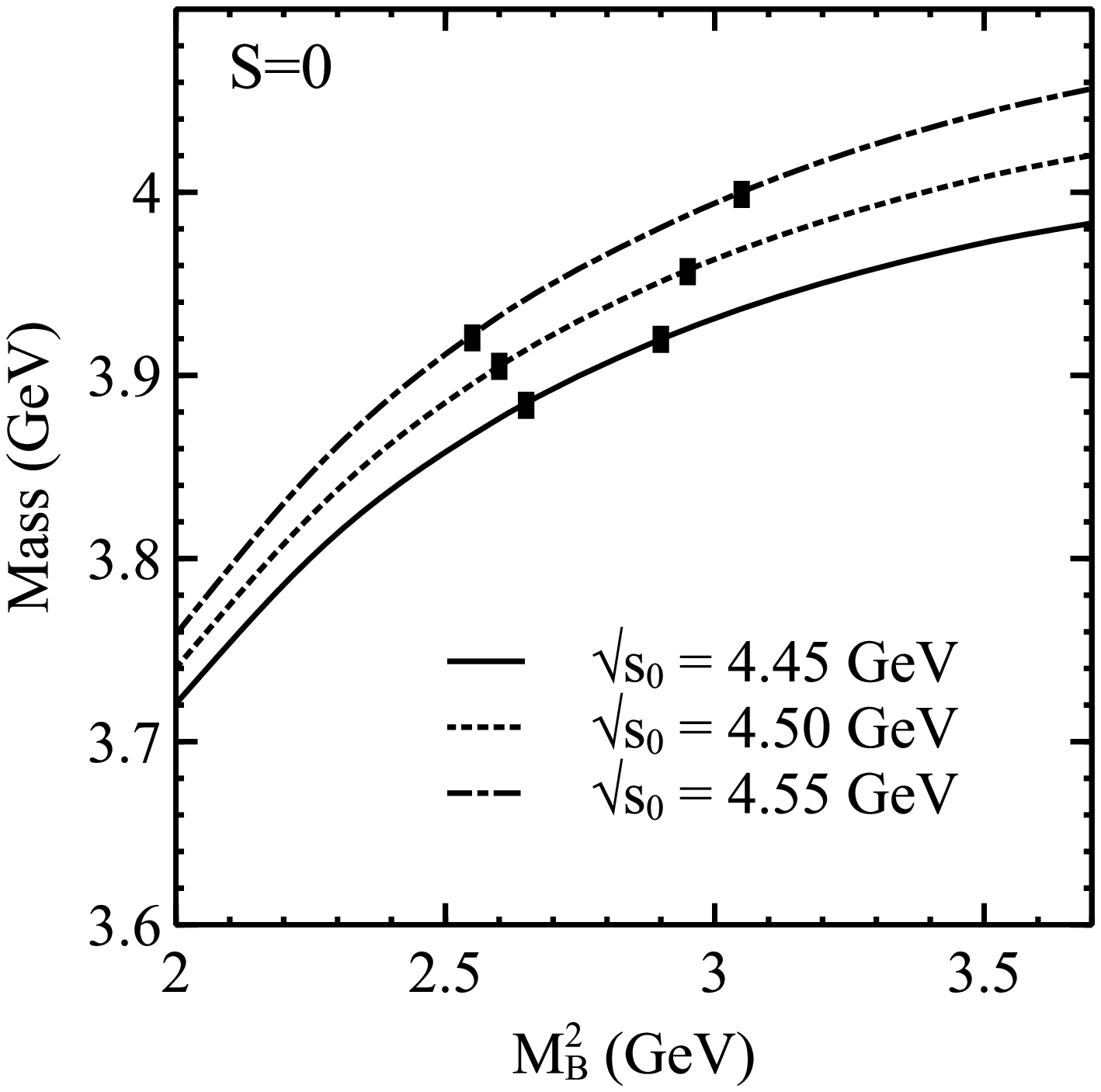} \includegraphics[width= 0.49\textwidth, height=8cm]{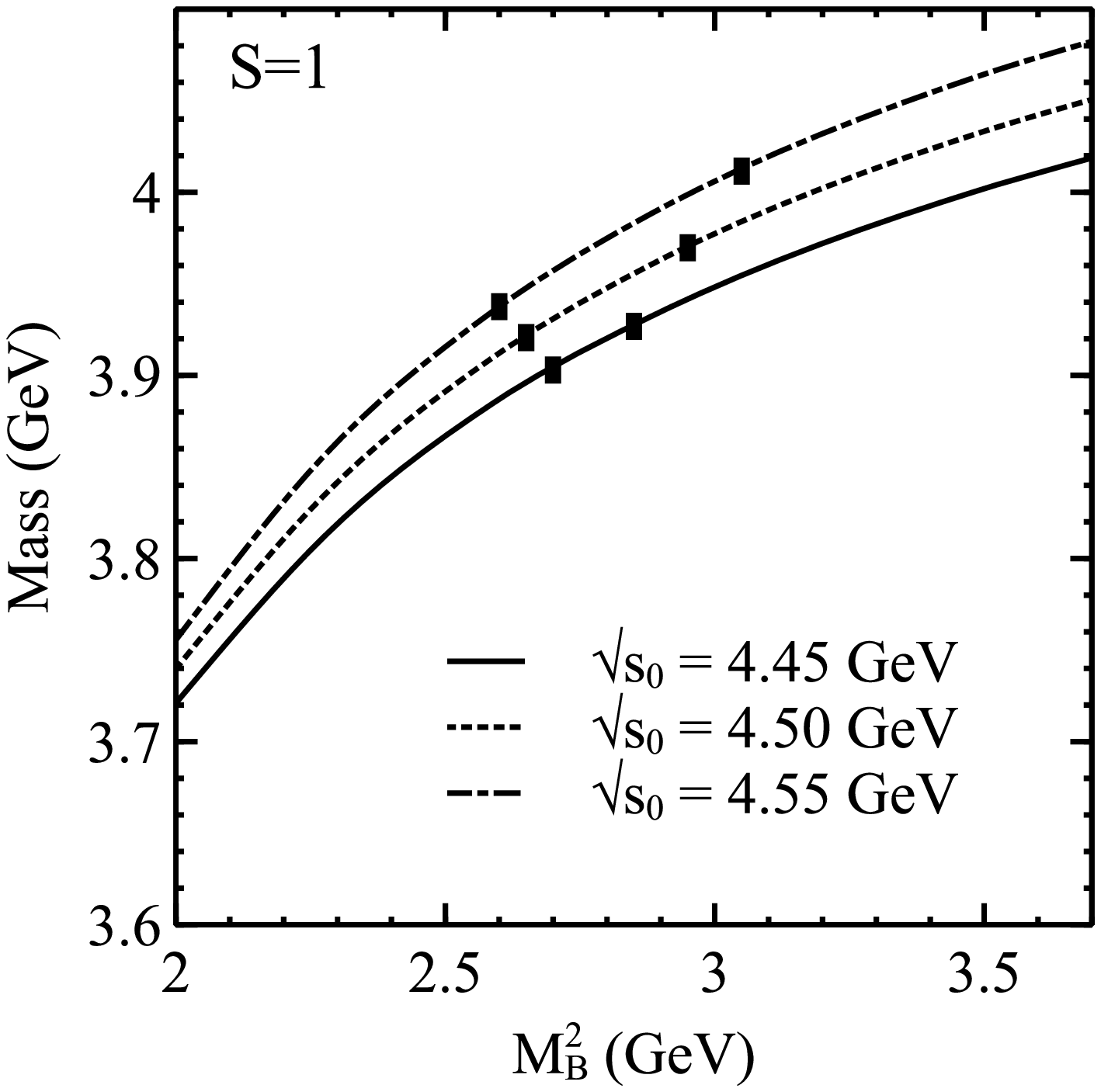}
\includegraphics[width= 0.49\textwidth, height=8cm]{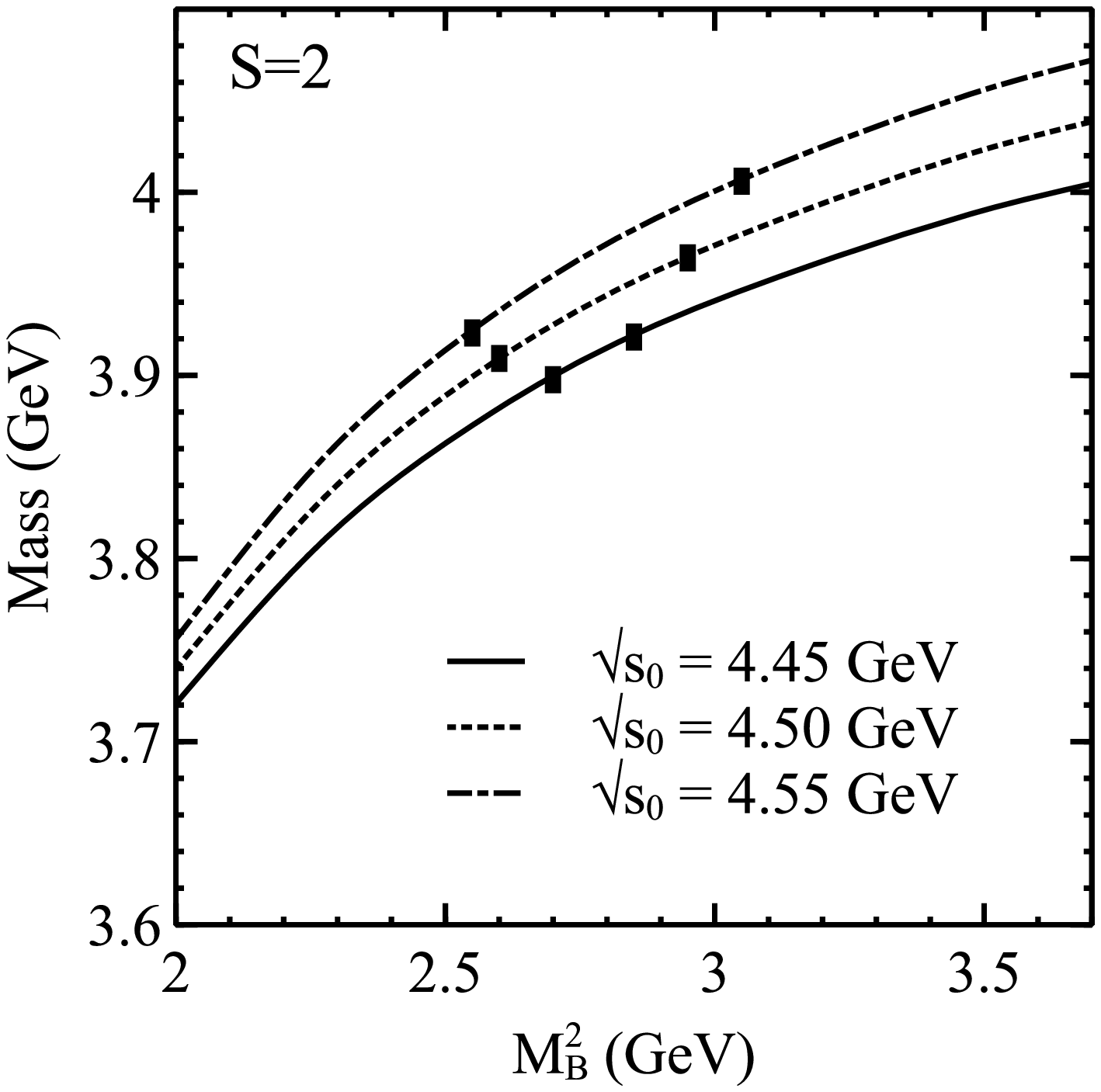}
\caption{Masses of the states obtained with spin-parity $0^+, 1^+$ and $2^+$. The meaning of $\sqrt{s}_0$, $S$ and $M_B^2$ here is same as in Fig.~\ref{polecont}. The valid Borel windows  are indicated in the figures by filled rectangles. }
\label{fig:mass}
\end{figure}

Further, in Fig.~\ref{fig:mass} we show the results obtained for the mass  in the three spin configurations for three different values of $\sqrt{s_0}$. The valid Borel window for the different cases is marked by a filled rectangle in the figure. 
These results show that there is a reasonable stability in the value of the mass of the states.  However, it is  further important to check the uncertainty present in our results due to the lack of precision with which the different condensates and the constituent mass of the charm quark are known. We thus vary all these inputs, one by one, and make an average over all the results obtained. The range within which the mass varies in the whole process gives us an idea of the total uncertainty present in our calculations. 
The masses obtained, considering all these sources of uncertainties,  in the  three cases are:
\begin{eqnarray}\label{res1}
M^{S=0} &=& \left(3943 \pm 104 \right)\,\, {\rm MeV},       \\\nonumber
M^{S=1} &=& \left(3950 \pm 105  \right)\,\,  {\rm MeV},        \\\nonumber
M^{S=2} &=& \left(3946 \pm 104 \right)\,\,{\rm MeV}.
\end{eqnarray}
 These results indicate that three states with $D^* \bar{D}^*$ structure and with very similar masses, same parity, but different spin exist. Within the error bars, one of these could  correspond to the recently observed $Z_c(4025)$  \cite{bes4025} for which the spin-parity is assumed to be $1^+$ in Ref.~\cite{bes4025}. The spin-parity conservation  excludes the $0^+$ assignment to this state since it has been found in the  $D^* \bar{D}^*$ mass spectrum in the $e^+ e^- \to \pi D^* \bar{D}^*$ process.  The $0^+$ state obtained here can probably be related with the $Z_1^+ (4050)$ state found in Ref.~\cite{belle4050_4250} in the $\pi \chi_{c1}$ system. A possibility of understanding $Z_1^+ (4050)$ as a $D^* \bar{D}^*$ molecule has earlier been investigated in some works \cite{Liu:2008tn,suhoung}. 

The $1^+$ and $2^+$ states  given in Eq.~(\ref{res1}) are both  compatible with the $Z_c(4025)$  found in the $D^* \bar{D}^*$ mass spectrum in Ref.~\cite{bes4025} although they could also correspond to  a state below the threshold (in agreement with Ref.~\cite{raquel2} as we will discuss shortly). The $2^+$ result  is in agreement with other work done using QCD sum rules with a tetraquark current~\cite{Qiao:2013dda}. The similarity in our results and those found in Ref.~\cite{Qiao:2013dda} are expected since it is well known that a tetraquark current and a molecular current with same quantum numbers can be related by a Fierz transformation \cite{Narison:2010pd}.  Our $1^+$ result is compatible with previous works where $D^* \bar{D}^*$  moleculelike currents were studied \cite{Chen:2013omd,Cui:2013vfa}, although the  currents in  Refs.~\cite{Chen:2013omd,Cui:2013vfa} were constructed  directly for an axial-vector state.
 
Resonance generation in the $D^* \bar{D}^*$ system  has been studied within effective field theories earlier \cite{Guo:2013sya,raquel2}.  In Refs.~\cite{Guo:2013sya} the $D^* \bar{D}^*$ system has been investigated with a formalism based on heavy quark symmetry and as a result an isospin 1 resonance with $J^\pi=1^+$  and mass ranging  between 3950-4017 MeV has been found  as  dynamically generated. This state has been associated to $Z_c(4025)$.  On the other hand, in Refs.~\cite{raquel2} a state with  only $J^\pi=2^+$, in the case of isospin 1, and mass 3900-3965 MeV has been found in the $D^* \bar{D}^*$ system.  As can be seen from Eq.~(\ref{res1}), our results in $1^+$ and $2^+$  configurations are both in good agreement with those found in Refs.~\cite{Guo:2013sya,raquel2}.

From our work, thus, both $1^+$ and $2^+$ spin-parity assignments seem to be plausible for  $Z_c(4025)$, but our mass results, within the error bars,  in the two cases are compatible with having a resonance or a bound state in the $D^* \bar{D}^*$ system.

We have also calculated the coupling of the states found in our work to the corresponding currents. We find
\begin{eqnarray}\label{res2}
\lambda^{S=0} &=&\left(17 \pm 4 \right)  \times 10^{-3}\,\,  {\rm GeV}^5,      \\\nonumber
\lambda^{S=1} &=&\left( 30 \pm 6 \right)  \times 10^{-3}\,\,  {\rm GeV}^5,      \\\nonumber
\lambda^{S=2} &=& \left(39 \pm 8 \right)  \times 10^{-3} \,\, {\rm GeV}^5, 
\end{eqnarray}
where the error bars have been obtained following the procedure mentioned earlier to calculate the uncertainty in the mass values.
Equation~(\ref{res2}) indicates that both currents with spin 1 and 2 couple  strongly to a $D^* \bar{D}^*$ state although the current-state coupling in
spin 2 seems to be slightly larger. From these results too, like in case of the masses found in Eq.~(\ref{res1}), both the spin-parity $1^+$ and $2^+$ seem to be equally associable  to  $Z_c(4025)$ \st{, if it exists}. In principle, with the results for the mass given in Eq.~(\ref{res1}), it is also not possible to exclude the possibility of associating the $1^+$ state found in our work  with $Z_c(3900)$. Although the $D^* \bar{D}^*$  threshold is about 120 MeV far from the mass of $Z_c(3900)$ and the  $D^* \bar{D}^*$ component should have a weight far too small in the wave function of $Z_c(3900)$, the coupling between the two can exist. Then if the $1^+$ state found here could be related to $Z_c(3900)$, we would be left with $2^+$ assignment for $Z_c(4025)$. 
Of course, within our formalism, it is not possible to clarify this issue and also if  the states found here are  resonances or bound states, which even questions the existence of $Z_c(4025)$. 
In fact very different explanations seem plausible for the signal found by the BES collaboration \cite{bes4025}  when the $D^* \bar{D}^*$ mass spectrum is calculated for the process $e^+ e^- \to \pi^\pm \left( D^* \bar{D}^*\right)^\pm$ considering  the production of  resonances/bound states with  spin-parities  $J^\pi = 1^+$ or $2^+$ in different partial waves \cite{usnew}.

\section{Summary} 
We have studied the $D^* \bar{D}^*$ system using QCD sum rules with the motivation to find if a moleculelike  
state could associated to the recently found $Z_c(4025)$ in the $\left(D^* \bar{D}^*\right)^\pm$ mass spectrum in the process: $e^+ e^- \to \pi^\pm \left(D^* \bar{D}^*\right)^\pm$ \cite{bes4025}.
The spin parity of this state is not known. The  $D^* \bar{D}^*$ system can have total spin 0, 1 or 2. We argue that the two mesons interact in s-wave since a very little energy is available to the $D^* \bar{D}^*$ system and thus such a state should have a positive parity. With this argument we obtain the correlation function corresponding to spin-parity $0^+, 1^+, 2^+$ using spin projectors of Ref.~\cite{us}.  With these correlation functions we carry out the calculation up to dimension 6 on the OPE side. We find the pole dominance and good OPE convergence for reasonable values of continuum thresholds. As a result we find three states with spin-parity $0^+, 1^+, 2^+$ each. The masses of all the states turn out to be very similar.  We conclude that  
both $1^+$ and $2^+$ assignments  could be associated with  $Z_c(4025)$. However, our results are also compatible with the formation of  bound states in line with the findings of Ref.~\cite{raquel2}.   The $0^+$ state found here can be related with  $Z_1^+ (4050)$~\cite{belle4050_4250}.  To conclude the article, we would like to say that it is very important to  obtain further experimental confirmation of the existence of these new states: $Z_c(4025)$ and $Z_c(4020)$.

\section{Acknowledgements} 
The authors would like to thank the Brazilian funding agencies FAPESP and CNPq for the financial support. The authors also acknowledge useful discussions with Profs. E. Oset and C. Z. Yuan.

\appendix*
\section{OPE results} \label{app}
In this section we give the results obtained for the different OPE terms in Eq.~(\ref{rho}). The superscript on $\rho$ in the following expressions denotes the spin of the current. Further, to write these results in a compact form we define following functions, 
\begin{eqnarray}
F_{\alpha_1, \alpha_2} &=& m_c^2 \left( \alpha_1 + \alpha_2 \right) - q^2 \alpha_1  \alpha_2\\\nonumber
g_{\alpha_1, \alpha_2} &=&  1- \alpha_1 - \alpha_2 \\\nonumber
h_{\alpha_1, \alpha_2} &=&  q^2  \alpha_1  \alpha_2 \\\nonumber
F_{\eta_1, \eta_2}&=& \frac{mc^2 \left(\eta_1 + \eta_2 \right)}{\eta_1 \eta_2}\\\nonumber
g_{\eta_1, \eta_2} &=& 1- \eta_1 - \eta_2,
\end{eqnarray}
where $\alpha_1, \alpha_2, \eta_1, \eta_2$ are variables of integration, $m_c$ is the constituent mass of the charm quark, $M_B$ is the Borel mass and $q$ is the running momentum in the 
correlation function.

\begin{eqnarray}\nonumber
\rho^{0}_{\textrm{pert}}&=&\frac{1}{\pi^6} \mathlarger{\int}\limits_{\alpha_{1min}}^{\alpha_{1max}} d\alpha_1 \mathlarger{\int}\limits_{\alpha_{2min}}^{\alpha_{2max}}d \alpha_2
 \left\{ \frac{g_{\alpha_1, \alpha_2}^3\left( \frac{F_{\alpha_1, \alpha_2}}{4} \left( F_{\alpha_1, \alpha_2} - 16 h_{\alpha_1, \alpha_2}\right) +  4 h_{\alpha_1, \alpha_2}^2\right)F_{\alpha_1, \alpha_2}^2}{2^{10} \alpha_1^3 \alpha_2^3 }   \right. \\\nonumber
&&\,\,\,\,\,\,+\left.   \frac{3 g_{\alpha_1, \alpha_2} F_{\alpha_1, \alpha_2}^4}{2^{11} \alpha_1^3 \alpha_2^3} -
\frac{3g_{\alpha_1, \alpha_2}^2 \left(\frac{F_{\alpha_1, \alpha_2}^4}{24} - \frac{1}{3}h_{\alpha_1, \alpha_2}F_{\alpha_1, \alpha_2}^3\right)}{2^{8} \alpha_1^3 \alpha_2^3} \right\}\\\nonumber
\rho^{1}_{\textrm{pert}}&=&\frac{1}{\pi^6} \mathlarger{\int}\limits_{\alpha_{1min}}^{\alpha_{1max}} d\alpha_1 \mathlarger{\int}\limits_{\alpha_{2min}}^{\alpha_{2max}}d \alpha_2
 \left\{ \frac{3 F_{\alpha_1, \alpha_2}^4 g_{\alpha_1, \alpha_2}}{2^{11}\alpha_1^3 \alpha_2^3} - \frac{9 g_{\alpha_1, \alpha_2}^2\left( \frac{F_{\alpha_1, \alpha_2}^4}{24} - \frac{F_{\alpha_1, \alpha_2}^3 h_{\alpha_1, \alpha_2}}{3}\right)}{2^8 \alpha_1^3 \alpha_2^3}  \right\}\\\nonumber
 \rho^{2}_{\textrm{pert}}&=&\frac{1}{\pi^6} \mathlarger{\int}\limits_{\alpha_{1min}}^{\alpha_{1max}} d\alpha_1 \mathlarger{\int}\limits_{\alpha_{2min}}^{\alpha_{2max}}d \alpha_2 \left\{  \frac{F_{\alpha_1, \alpha_2}^2 g_{\alpha_1, \alpha_2}^3 \left[\frac{F_{\alpha_1, \alpha_2}}{4} \left( F_{\alpha_1, \alpha_2} - 16 h_{\alpha_1, \alpha_2}\right) + 4 h_{\alpha_1, \alpha_2}^2\right]}{2^9 \alpha_1^3 \alpha_2^3}\right.\\\nonumber
&&+ \frac{3 F_{\alpha_1, \alpha_2}^4 g_{\alpha_1, \alpha_2}}{2^{11}\alpha_1^3 \alpha_2^3} + \frac{3 g_{\alpha_1, \alpha_2}^2\left(\frac{F_{\alpha_1, \alpha_2}^4}{24}-\frac{F_{\alpha_1, \alpha_2}^3 h_{\alpha_1, \alpha_2}}{3}\right)}{2^8 \alpha_1^3 \alpha_2^3}
\end{eqnarray}

\begin{eqnarray}\nonumber
\rho^{0}_{\langle\bar q q\rangle}&=& -\frac{ m_c \langle\bar q q\rangle}{2^7 \pi^4}\mathlarger{\int}\limits_{\alpha_{1min}}^{\alpha_{1max}} d\alpha_1 \mathlarger{\int}\limits_{\alpha_{2min}}^{\alpha_{2max}}d \alpha_2   \left(\frac{\alpha_1 + \alpha_2}{\alpha_1^2 \alpha_2^2} \right)\left(g_{\alpha_1, \alpha_2}\left[F_{\alpha_1, \alpha_2}^2 - 4 F_{\alpha_1, \alpha_2}h_{\alpha_1, \alpha_2}\right]  + F_{\alpha_1, \alpha_2}^2\right)\\\nonumber
\rho^{1}_{\langle\bar q q\rangle}&=& -3\rho^{0}_{\langle\bar q q\rangle}\\\nonumber
\rho^{2}_{\langle\bar q q\rangle}&=& 5\rho^{0}_{\langle\bar q q\rangle}
\end{eqnarray}

\begin{eqnarray}\nonumber
\rho^{0}_{\langle g^2 G^2\rangle}&=& \frac{\langle g^2G^2 \rangle }{\pi^6} \mathlarger{\int}\limits_{\alpha_{1min}}^{\alpha_{1max}} d\alpha_1 \mathlarger{\int}\limits_{\alpha_{2min}}^{\alpha_{2max}}d \alpha_2 \Biggl\{ 
\left(-\frac{\alpha_1+\alpha_2}{24 \alpha_1^2 \alpha_2^2}\right)\left[\frac{g_{\alpha_1, \alpha_2}}{2^6}\left(\frac{ F_{\alpha_1, \alpha_2}^2}{2} - 2 F_{\alpha_1, \alpha_2} h_{\alpha_1, \alpha_2}\right)  \right. \Biggr.\\\nonumber
&&\left.  -\frac{F_{\alpha_1, \alpha_2}^2}{2^9}  -\frac{g_{\alpha_1, \alpha_2}^2\left(\frac{3}{2}  F_{\alpha_1, \alpha_2} \left( F_{\alpha_1, \alpha_2} -8 h_{\alpha_1, \alpha_2} \right) + 4 h_{\alpha_1, \alpha_2}^2 \right)}{2^9} \right] 
+ m_c^2\left( \frac{\alpha_1^3 + \alpha_2^3}{\alpha_1^3 \alpha_2^3} \right)\\\nonumber
&&\Biggl.
\times \left[ -\frac{g_{\alpha_1, \alpha_2}^3 \left( 4 h_{\alpha_1, \alpha_2} - F_{\alpha_1, \alpha_2} \right)}{3\cdot2^{12}} + \frac{F_{\alpha_1, \alpha_2} g_{\alpha_1, \alpha_2} }{2^{11}} -
 \frac{g_{\alpha_1, \alpha_2}^2 \left (m_c^2 \left(\alpha_1 + \alpha_2 \right) - 3 h_{\alpha_1, \alpha_2} \right)}{3 \cdot 2^{11}}\right]
\Biggr\}\\\nonumber
&-&\frac{m_c^6 \langle g^2G^2 \rangle }{9\cdot2^{10}\pi^6}\mathlarger{\int}\limits_{\eta_{1min}}^{\eta_{1max}} d\eta_1 \mathlarger{\int}\limits_{\eta_{2min}}^{\eta_{2max}}d \eta_2 
\frac{g_{\eta_1, \eta_2}^3 \left(\eta_1+ \eta_2\right)^2  \left(\eta_1^3+ \eta_2^3\right)}{\eta_1^4 \eta_2^4} \delta \left(s - F_{\eta_1, \eta_2}  \right)
\end{eqnarray}

\begin{eqnarray}\nonumber
\rho^{1}_{\langle g^2 G^2\rangle}&=& \frac{\langle g^2G^2 \rangle }{\pi^6}  \mathlarger{\int}\limits_{\alpha_{1min}}^{\alpha_{1max}} d\alpha_1 \mathlarger{\int}\limits_{\alpha_{2min}}^{\alpha_{2max}}d \alpha_2 \Biggl\{ \left(\frac{\alpha_1+\alpha_2}{ \alpha_1^2 \alpha_2^2}\right)  \left(\frac{3 F_{\alpha_1, \alpha_2}^2 }{2^{12}} \right)-\frac{m_c^2}{2^{11}}\left( \frac{\alpha_1^3 + \alpha_2^3}{\alpha_1^3 \alpha_2^3} \right) \Biggr. \\\nonumber
&& \Biggl. \times  \left( g_{\alpha_1, \alpha_2}^2 \left[ m_c^2 \left( \alpha_1+\alpha_2 \right) - 3 h_{\alpha_1, \alpha_2}\right] - F_{\alpha_1, \alpha_2} g_{\alpha_1, \alpha_2}\right) \Biggr\}\\\nonumber
\rho^{2}_{\langle g^2 G^2\rangle}&=& \frac{\langle g^2G^2 \rangle }{\pi^6}  \mathlarger{\int}\limits_{\alpha_{1min}}^{\alpha_{1max}} d\alpha_1 \mathlarger{\int}\limits_{\alpha_{2min}}^{\alpha_{2max}}d \alpha_2 \Biggl\{ \left(\frac{\alpha_1+\alpha_2}{24 \alpha_1^2 \alpha_2^2}\right)\left[-\frac{g_{\alpha_1, \alpha_2}}{2^5}\left(\frac{ F_{\alpha_1, \alpha_2}^2}{2} - 2 F_{\alpha_1, \alpha_2} h_{\alpha_1, \alpha_2}\right)  \Biggr.\right.
\\\nonumber
&&\left.  -\frac{7 F_{\alpha_1, \alpha_2}^2}{2^9}  +\frac{g_{\alpha_1, \alpha_2}^2\left(\frac{3}{2}  F_{\alpha_1, \alpha_2} \left( F_{\alpha_1, \alpha_2} -8 h_{\alpha_1, \alpha_2} \right) + 4 h_{\alpha_1, \alpha_2}^2 \right)}{2^8} \right] 
+ m_c^2\left( \frac{\alpha_1^3 + \alpha_2^3}{\alpha_1^3 \alpha_2^3} \right)\\\nonumber
&&\Biggl.
\times \left[ -\frac{g_{\alpha_1, \alpha_2}^3 \left( 4 h_{\alpha_1, \alpha_2} - F_{\alpha_1, \alpha_2} \right)}{3\cdot2^{11}} + \frac{F_{\alpha_1, \alpha_2} g_{\alpha_1, \alpha_2} }{2^{11}} +
 \frac{g_{\alpha_1, \alpha_2}^2 \left (m_c^2 \left(\alpha_1 + \alpha_2 \right) - 3 h_{\alpha_1, \alpha_2} \right)}{3 \cdot 2^{11}}\right]
\Biggr\}\\\nonumber
&-&\frac{m_c^6 \langle g^2G^2 \rangle }{9\cdot2^{9}\pi^6}\mathlarger{\int}\limits_{\eta_{1min}}^{\eta_{1max}} d\eta_1 \mathlarger{\int}\limits_{\eta_{2min}}^{\eta_{2max}}d \eta_2 
\frac{g_{\eta_1, \eta_2}^3 \left(\eta_1+ \eta_2\right)^2  \left(\eta_1^3+ \eta_2^3\right)}{\eta_1^4 \eta_2^4} \delta \left(s - F_{\eta_1, \eta_2}  \right),
\end{eqnarray}
where the variable $s = q^2$.

\begin{adjustwidth}{-5cm}{}
\begin{eqnarray}\nonumber
\rho^{0}_{\langle\bar q g\sigma G q\rangle}&=& \frac{-m_c \langle\bar q g\sigma G q\rangle}{2^8 \pi^4} \mathlarger{\int}\limits_{\alpha_{1min}}^{\alpha_{1max}} \frac{d\alpha_1}{\alpha_1}
 \Biggl\{\frac{\left(m_c^2 - s \alpha_1 \left(1- \alpha_1\right)\right)}{\left(1- \alpha_1\right)} \Biggr. \\\nonumber
&+&
\mathlarger{\int}\limits_{\alpha_{2min}}^{\alpha_{2max}} \left.\frac{d \alpha_2}{\alpha_2} \left( m_c^2\left( \alpha_1 + \alpha_2 \right) - 3 h_{\alpha_1, \alpha_2}\right) \left( \alpha_1+  \alpha_2 \right) \right\}
\end{eqnarray}
 \end{adjustwidth}
 \begin{adjustwidth}{-11cm}{}
\begin{eqnarray}\nonumber
\rho^{1}_{\langle\bar q g\sigma G q\rangle}&=&-3 \rho^{0}_{\langle\bar q g\sigma G q\rangle}\\\nonumber
\rho^{2}_{\langle\bar q g\sigma G q\rangle}&=&5 \rho^{0}_{\langle\bar q g\sigma G q\rangle}
 \end{eqnarray}
 \end{adjustwidth}

 \begin{adjustwidth}{-11cm}{} 
\begin{eqnarray}\nonumber
\rho^{0}_{{\langle\bar q q\rangle}^2}&=&\mathlarger{\int}\limits_{\alpha_{1min}}^{\alpha_{1max}} d\alpha_1 \frac{m_c^2 \langle\bar q q\rangle^2}{2^4 \pi^2} \\\nonumber
\rho^{1}_{{\langle\bar q q\rangle}^2}&=&-3 \rho^{0}_{{\langle\bar q q\rangle}^2}\\\nonumber
\rho^{2}_{{\langle\bar q q\rangle}^2}&=&5\rho^{0}_{{\langle\bar q q\rangle}^2}
\end{eqnarray}
\end{adjustwidth}

\begin{eqnarray}\nonumber
\rho^{0}_{\langle g^3 G^3\rangle}&=& \frac{\langle g^3G^3 \rangle }{\pi^6}\mathlarger{\int}\limits_{\alpha_{1min}}^{\alpha_{1max}} d\alpha_1 \mathlarger{\int}\limits_{\alpha_{2min}}^{\alpha_{2max}}d \alpha_2
 \left\{ \left(\frac{\alpha_1^3+\alpha_2^3}{ \alpha_1^3 \alpha_2^3}\right)\left[-\frac{g_{\alpha_1, \alpha_2}^3\left(4 h_{\alpha_1, \alpha_2} -  F_{\alpha_1, \alpha_2} \right)}{3\cdot2^{14}} 
  +\frac{F_{\alpha_1, \alpha_2}g_{\alpha_1, \alpha_2}}{2^{13}}
  \right. \right.\\\nonumber
&&\left.  \left. -\frac{g_{\alpha_1, \alpha_2}^2\left(m_c^2 \left( \alpha_1+ \alpha_2 \right) - 3 h_{\alpha_1, \alpha_2} \right)}{3\cdot2^{13}} \right] 
+ 2 m_c^2\left( \frac{\alpha_1^4 + \alpha_2^4}{\alpha_1^3 \alpha_2^3} \right)
 \left[ \frac{g_{\alpha_1, \alpha_2}^3}{3\cdot2^{14}} - \frac{g_{\alpha_1, \alpha_2}^2 }{3\cdot2^{13}} + \frac{g_{\alpha_1, \alpha_2}}{2^{13}}\right]
\right\}\\\nonumber
&&+\frac{m_c^4 \langle g^3G^3 \rangle }{3\cdot2^{12}\pi^6}\mathlarger{\int}\limits_{\eta_{1min}}^{\eta_{1max}} d\eta_1 \mathlarger{\int}\limits_{\eta_{2min}}^{\eta_{2max}}d \eta_2 \left(\frac{ g_{\eta_1, \eta_2}^2\left(\eta_1+ \eta_2\right)}{\eta_1^4 \eta_2^4}\right) \left[-\frac{g_{\eta_1, \eta_2}\left(\eta_1^3+ \eta_2^3\right)\left(\eta_1+ \eta_2 \right)}{3} \right.\\\nonumber
&&+ \left. 2\left(\frac{g_{\eta_1, \eta_2}}{3}\left[\frac{F_{\eta_1, \eta_2}}{M_B^2} + 1 \right]-1\right)\left(\eta_1^4 + \eta_2^4\right) \right]
 \delta \left(s - F_{\eta_1, \eta_2} \right)\\\nonumber
 \rho^{1}_{\langle g^3 G^3\rangle}&=& \frac{\langle g^3G^3 \rangle }{\pi^6}\mathlarger{\int}\limits_{\alpha_{1min}}^{\alpha_{1max}} d\alpha_1 \mathlarger{\int}\limits_{\alpha_{2min}}^{\alpha_{2max}}d \alpha_2  \left\{ \left(\frac{\alpha_1^3+\alpha_2^3}{ \alpha_1^3 \alpha_2^3}\right)\left[
 -\frac{g_{\alpha_1, \alpha_2}^2\left(m_c^2 \left( \alpha_1+ \alpha_2 \right) - 3 h_{\alpha_1, \alpha_2} \right)}{2^{13}} 
  \right. \right.\\\nonumber
  &&+ \left. \left. \frac{F_{\alpha_1, \alpha_2}g_{\alpha_1, \alpha_2}}{2^{13}} \right]+ m_c^2 \left( \frac{\alpha_1^4 + \alpha_2^4}{\alpha_1^3 \alpha_2^3} \right)\left( \frac{g_{\alpha_1, \alpha_2}}{2^{12}} - \frac{g_{\alpha_1, \alpha_2}^2 }{2^{12}} \right) \right\}  \\\nonumber
  &&- \frac{m_c^4 \langle g^3G^3 \rangle }{2^{11}\pi^6}\mathlarger{\int}\limits_{\eta_{1min}}^{\eta_{1max}} d\eta_1 \mathlarger{\int}\limits_{\eta_{2min}}^{\eta_{2max}}d \eta_2
  \left(\frac{ g_{\eta_1, \eta_2}^2\left(\eta_1+ \eta_2\right)\left(\eta_1^4+ \eta_2^4\right)}{\eta_1^4 \eta_2^4}\right) \delta \left(s - F_{\eta_1, \eta_2} \right)
 \end{eqnarray}

\begin{eqnarray}\nonumber
  \rho^{2}_{\langle g^3 G^3\rangle}&=& \frac{\langle g^3G^3 \rangle }{\pi^6}\mathlarger{\int}\limits_{\alpha_{1min}}^{\alpha_{1max}} d\alpha_1 \mathlarger{\int}\limits_{\alpha_{2min}}^{\alpha_{2max}}d \alpha_2
  \left\{ \left(\frac{\alpha_1^3+\alpha_2^3}{ \alpha_1^3 \alpha_2^3}\right)\left[-\frac{g_{\alpha_1, \alpha_2}^3\left(4 h_{\alpha_1, \alpha_2} -  F_{\alpha_1, \alpha_2} \right)}{3\cdot2^{13}} 
  +\frac{F_{\alpha_1, \alpha_2}g_{\alpha_1, \alpha_2}}{2^{13}}
  \right. \right.\\\nonumber
  &&\left.  \left. +\frac{g_{\alpha_1, \alpha_2}^2\left(m_c^2 \left( \alpha_1+ \alpha_2 \right) - 3 h_{\alpha_1, \alpha_2} \right)}{3\cdot2^{13}} \right] 
+ 2 m_c^2\left( \frac{\alpha_1^4 + \alpha_2^4}{\alpha_1^3 \alpha_2^3} \right)
 \left[ \frac{g_{\alpha_1, \alpha_2}^3}{3\cdot2^{13}} + \frac{g_{\alpha_1, \alpha_2}^2 }{3\cdot2^{13}} + \frac{g_{\alpha_1, \alpha_2}}{2^{13}}\right]
\right\}\\\nonumber
&&+\frac{m_c^4 \langle g^3G^3 \rangle }{3\cdot2^{11}\pi^6}\mathlarger{\int}\limits_{\eta_{1min}}^{\eta_{1max}} d\eta_1 \mathlarger{\int}\limits_{\eta_{2min}}^{\eta_{2max}}d \eta_2 
\left(\frac{ g_{\eta_1, \eta_2}^2\left(\eta_1+ \eta_2\right)}{\eta_1^4 \eta_2^4}\right) \left[-\frac{g_{\eta_1, \eta_2}\left(\eta_1^3+ \eta_2^3\right)\left(\eta_1+ \eta_2 \right)}{3} \right.\\\nonumber
&&+ \left. 2\left(\frac{g_{\eta_1, \eta_2}}{3}\left[\frac{F_{\eta_1, \eta_2}}{M_B^2} + 1 \right]+\frac{1}{2}\right)\left(\eta_1^4 + \eta_2^4\right) \right]
 \delta \left(s - F_{\eta_1, \eta_2} \right)
\end{eqnarray}

The limits of integration in above expressions are
\begin{eqnarray}
\alpha_{1min} &=&\frac{ 1- \sqrt{1 - \frac{4 m_c^2}{q^2}}}{2}, \hspace{0.2cm} \alpha_{1max} =\frac{ 1+ \sqrt{1 - \frac{4 m_c^2}{q^2}}}{2}, \nonumber
\end{eqnarray}
\begin{eqnarray}
\alpha_{2min} &=&\frac{ m_c^2 \alpha_1}{\left(\alpha_1 q^2- m_c^2\right)}, \hspace{0.2cm} \alpha_{2max} = 1 - \alpha_1,\nonumber
\end{eqnarray}
\begin{eqnarray}
\eta_{1min} &=& 0, \hspace{0.2cm}\eta_{1max} = 1\nonumber
\end{eqnarray}
\begin{eqnarray}
\eta_{2min} &=& 0, \hspace{0.2cm}\eta_{2max} = 1 - \eta_1. \nonumber
\end{eqnarray}


\begin{thebibliography}{99}

\bibitem{bes4025} 
  M.~Ablikim {\it et al.}  [BESIII Collaboration],
  arXiv:1308.2760 [hep-ex].

\bibitem{bes4025_2} 
  M.~Ablikim {\it et al.}  [ BESIII Collaboration],
  arXiv:1309.1896 [hep-ex].

\bibitem{belle4430_1} 
  S.~K.~Choi {\it et al.}  [BELLE Collaboration],
  Phys.\ Rev.\ Lett.\  {\bf 100}, 142001 (2008)
  [arXiv:0708.1790 [hep-ex]].

\bibitem{belle4430_2} 
  R.~Mizuk {\it et al.}  [BELLE Collaboration],
  Phys.\ Rev.\ D {\bf 80}, 031104 (2009)
  [arXiv:0905.2869 [hep-ex]].

\bibitem{belle4430_3} 
  K.~Chilikin {\it et al.}  [Belle Collaboration],
  arXiv:1306.4894 [hep-ex].


\bibitem{belle4050_4250} 
  R.~Mizuk {\it et al.}  [Belle Collaboration],
  Phys.\ Rev.\ D {\bf 78}, 072004 (2008)
  [arXiv:0806.4098 [hep-ex]].

\bibitem{bes3900} 
  M.~Ablikim {\it et al.}  [BESIII Collaboration],
  Phys.\ Rev.\ Lett.\  {\bf 110}, 252001 (2013)
  [arXiv:1303.5949 [hep-ex]].
  
\bibitem{belle3900} 
  Z.~Q.~Liu {\it et al.}  [Belle Collaboration],
  Phys.\ Rev.\ Lett.\  {\bf 110}, 252002 (2013)
  [arXiv:1304.0121 [hep-ex]].

\bibitem{babar4430} 
  B.~Aubert {\it et al.}  [BaBar Collaboration],
  Phys.\ Rev.\ D {\bf 79}, 112001 (2009)
  [arXiv:0811.0564 [hep-ex]].
  
  
\bibitem{babar4050_4250} 
  J.~P.~Lees {\it et al.}  [BaBar Collaboration],
  Phys.\ Rev.\ D {\bf 85}, 052003 (2012)
  [arXiv:1111.5919 [hep-ex]].

  

\bibitem{us} 
  A.~Mart\'inez~Torres, K.~P.~Khemchandani, M.~Nielsen, F.~S.~Navarra and E.~Oset,
  arXiv:1307.1724 [nucl-th].
  
  
\bibitem{Chen:2013omd} 
  W.~Chen, T.~G.~Steele, M.~-L.~Du and S.~-L.~Zhu,
  arXiv:1308.5060 [hep-ph].
  
  
\bibitem{Cui:2013vfa} 
  C.~-Y.~Cui, Y.~-L.~Liu and M.~-Q.~Huang,
  arXiv:1308.3625 [hep-ph].
  
\bibitem{Qiao:2013dda} 
  C.~-F.~Qiao and L.~Tang,
  arXiv:1308.3439 [hep-ph].
  
\bibitem{He:2013nwa} 
  J.~He, X.~Liu, Z.~-F.~Sun and S.~-L.~Zhu,
  arXiv:1308.2999 [hep-ph].

\bibitem{Wang:2013qwa} 
  X.~Wang, Y.~Sun, D.~-Y.~Chen, X.~Liu and T.~Matsuki,
  arXiv:1308.3158 [hep-ph].

  
\bibitem{Guo:2013sya} 
  F.~-K.~Guo, C.~Hidalgo-Duque, J.~Nieves and M.~P.~Valderrama,
  arXiv:1303.6608 [hep-ph].
  
\bibitem{raquel2} 
  R.~Molina and E.~Oset,
  Phys.\ Rev.\ D {\bf 80}, 114013 (2009)
  [arXiv:0907.3043 [hep-ph]].
  
   \bibitem{raquel}
 R.~Molina, H.~Nagahiro, A.~Hosaka and E.~Oset,
  Phys.\ Rev.\ D {\bf 80}, 014025 (2009).
  
 \bibitem{project}
  J.~Govaerts, L.~J.~Reinders, P.~Francken, X.~Gonze and J.~Weyers,
  Nucl.\ Phys.\ B {\bf 284}, 674 (1987).

\bibitem{svz}
M.A. Shifman, A.I. and Vainshtein and V.I. Zakharov,
Nucl. Phys. B {\bf 147}, 385 (1979);

\bibitem{reviews1}
 P.~Colangelo and A.~Khodjamirian,
  In *Shifman, M. (ed.): At the frontier of particle physics, vol. 3* 1495-1576
  [hep-ph/0010175].

 \bibitem{reviews2}
S. Narison, {\it QCD as a theory of hadrons,
Cambridge Monogr. Part. Phys. Nucl. Phys. Cosmol.} {\bf 17}, 1 (2002); {\it QCD spectral sum rules ,  World Sci. Lect. Notes Phys.} 
{\bf 26}, 1 (1989);
{ Acta Phys. Pol.} B {\bf 26}, 687 (1995); { Riv. Nuov. Cim.} {\bf 10N2}, 1
(1987); { Phys. Rept.} {\bf 84}, 263 (1982).

 \bibitem{reviews3} S.~Narison, Phys.\ Lett.\ B {\bf 216}, 191 (1989); {\bf 341}, 73 
(1994); {\bf 361}, 121 (1995), {\bf 387}, 162 (1996); {\bf 466}, 345 (1999); 
{\bf 624}, 223 (2005).

\bibitem{reviews4}
 M.~Nielsen, F.~S.~Navarra and S.~H.~Lee,
  Phys.\ Rept.\  {\bf 497}, 41 (2010).

\bibitem{our1}
  A.~Martinez Torres, K.~P.~Khemchandani, M.~Nielsen and F.~S.~Navarra,
  Phys.\ Rev.\ D {\bf 87}, 034025 (2013)
  

  
  \bibitem{Liu:2008tn} 
  X.~Liu, Z.~-G.~Luo, Y.~-R.~Liu and S.~-L.~Zhu,
  Eur.\ Phys.\ J.\ C {\bf 61}, 411 (2009)
  [arXiv:0808.0073 [hep-ph]].
  

  
\bibitem{suhoung} 
  S.~H.~Lee, K.~Morita and M.~Nielsen,
  Nucl.\ Phys.\ A {\bf 815}, 29 (2009)
  [arXiv:0808.0690 [hep-ph]].
  
\bibitem{Narison:2010pd} 
  S.~Narison, F.~S.~Navarra and M.~Nielsen,
  Phys.\ Rev.\ D {\bf 83}, 016004 (2011)
  [arXiv:1006.4802 [hep-ph]].
  
 \bibitem{usnew}
  A.~Martinez.~Torres, K.~P.~Khemchandani, F.~S.~Navarra, M.~Nielsen and E.~Oset,
  arXiv:1310.1119 [hep-ph].
 
 
 
\bibitem{rafael} 
  R.~M.~Albuquerque and M.~Nielsen,
  Nucl.\ Phys.\ A {\bf 815}, 53 (2009)
  [Erratum-ibid.\ A {\bf 857}, 48 (2011)]
  [arXiv:0804.4817 [hep-ph], arXiv:1104.2192 [hep-ph]].
  
 \end{thebibliography}
\end{document}